\documentclass[11pt,a4paper]{article}
\pdfoutput=1
\usepackage{jheppub}
\usepackage{subcaption}
\usepackage{tikz}

\DeclareMathOperator{\Tr}{Tr}

\newcommand{\ri}{\mathrm{i}}

\newcommand{\cob}{\delta}

\newcommand{\hf}{\frac{1}{2}}

\newcommand{\del}{\partial}
\newcommand{\wg}{\wedge}
\newcommand{\lap}{\Delta}
\newcommand{\bra}{\langle}
\newcommand{\ket}{\rangle}
\newcommand{\la}{\lambda}

\newcommand{\ka}{\kappa}
\newcommand{\h}[1]{\widehat{#1}}

\newcommand{\rt}[1]{\sqrt{#1}}

\begin{document}

\title{Connected correlator of 1/2 BPS Wilson loops in $\mathcal{N}=4$ SYM}
\author{Kazumi Okuyama}
\affiliation{Department of Physics, 
Shinshu University, Matsumoto 390-8621, Japan}

\emailAdd{kazumi@azusa.shinshu-u.ac.jp}

\abstract{
We study the 
connected correlator of 1/2 BPS winding Wilson loops in $\mathcal{N}=4$ 
$U(N)$ super Yang-Mills theory, where
those Wilson loops are on top of each other along the same circle.
We find the exact finite $N$
expression of the connected correlator of such Wilson loops.
We show that the $1/N$ expansion of this exact result is reproduced from
the topological recursion of Gaussian matrix model.
We also study the exact finite $N$ expression of 
the generating function of 
1/2 BPS Wilson loops
in the symmetric representation.
}

\maketitle
\section{Introduction}
1/2 BPS circular Wilson loops in 4d $\mathcal{N}=4$ $U(N)$ super Yang-Mills (SYM)
serve as a useful testing ground for
\textit{precision holography} between $\mathcal{N}=4$ SYM and 
the type IIB string theory on $AdS_5\times S^5$, since their expectation values can be
computed exactly by a Gaussian matrix model
at finite $N$ with arbitrary value of the 
't Hooft coupling $\la$ \cite{Erickson:2000af,Drukker:2000rr,Pestun:2007rz}. 

At the leading order in the large $N$ limit, the expectation value of
the 1/2 BPS Wilson loops are shown to agree with the on-shell action of fundamental string,
D3-brane, or D5-brane for the Wilson loops in the 
fundamental, symmetric, or anti-symmetric representations, respectively
\cite{Maldacena:1998im,Rey:1998ik,Drukker:2005kx,Yamaguchi:2006tq,Hartnoll:2006is,Okuyama:2006jc,Gomis:2006sb,Gomis:2006im}
(see also \cite{Zarembo:2016bbk} for a review).
We can even go further to study the $1/N$ corrections to these
cases, but the status of the matching at the subleading order is still unsettled
\cite{Faraggi:2014tna,Buchbinder:2014nia,Faraggi:2011ge,Faraggi:2011bb,Chen-Lin:2016kkk,Gordon:2017dvy}.
One can also consider the Wilson loops in very large representations
where the backreaction to the bulk geometry is so large
and the dual geometry is no longer a pure $AdS_5$ but it is replaced by a bubbling geometry
with non-trivial topology \cite{Yamaguchi:2006te,Lunin:2006xr,DHoker:2007mci,Okuda:2008px}.
We hope that one can also study such effect of topology change 
in the regime of quantum gravity using the exact result of 1/2 BPS Wilson loops.   

To explore the behavior of 1/2 BPS Wilson loops with various size
of representations, it is important to evaluate
the Gaussian matrix integral in a closed form for arbitrary 
representations.
This problem was partly solved in 
\cite{Fiol:2013hna}, but the result for general representation still involves
a complicated summation over partitions.
In the case of anti-symmetric representations, 
their generating function has a rather simple form and we can study
its behavior either analytically or numerically.
This study was initiated in \cite{Horikoshi:2016hds} and
the $1/N$ corrections of anti-symmetric Wilson loops are further studied in 
\cite{Gordon:2017dvy,Okuyama:2017feo,CanazasGaray:2018cpk}. 

In this paper, we point out that the connected correlator of 
winding Wilson loops
has a simple expression at finite $N$,
where the Wilson loops in question are on top
of each other along the same circle.\footnote{
When Wilson loops are separated in spacetime, 
there is a phase transition at some critical value of the distance between
loops \cite{Olesen:2000ji,Gross:1998gk}. We will not consider such cases in this paper.
}
This is expected since the
natural object to compute in the matrix model 
is not the trace of matrix in a irreducible representation but 
the correlator of resolvents with fixed number of holes and handles.
We also show that the $1/N$ expansion of the exact result of 
connected correlator of winding Wilson loops
is correctly reproduced from the 
topological recursion of Gaussian matrix model\footnote{The genus expansion of Wilson loop correlator
was also studied in \cite{Akemann:2001st}.},
as demonstrated in \cite{Okuyama:2017feo} in the case of anti-symmetric representation.
We also write down the generating function of 1/2 BPS Wilson loops in the symmetric representations.
We find an interesting relation between the generating functions
of the symmetric and the anti-symmetric Wilson loops. 

This paper is organized as follows.
In section \ref{sec:exact}, we write down the
exact expression of the connected correlator of 1/2 BPS winding Wilson loops at finite $N$.
In section \ref{sec:top}, we show that the $1/N$ corrections to such connected correlators are
systematically obtained by the topological recursion of Gaussian matrix model.
In section \ref{sec:comparison}, we compare the exact result of connected correlator at finite
$N$ and the analytic result of $1/N$ expansion obtained from the topological recursion, and
we find perfect match for both small $\la$ and finite $\la$ regimes. 
In section \ref{sec:large}, we comment on the limit of large winding number and
the bulk D3-brane picture of connected correlators.
In section \ref{sec:sym}, we write down the
exact generating function of 1/2 BPS Wilson loops in the symmetric representation and 
comment on its relation to the generating function of anti-symmetric representation.
We conclude in section \ref{sec:conclusion} with a brief discussion on the future
problems. 
\section{Connected correlator of winding Wilson loops at finite $N$ \label{sec:exact}}

As shown in \cite{Erickson:2000af,Drukker:2000rr,Pestun:2007rz},
the expectation value of 1/2 BPS circular Wilson loop in 
4d $\mathcal{N}=4$ $U(N)$ SYM 
is exactly given by a Gaussian matrix model.
The expectation value of the Wilson loop in the representation $R$
 is written as
\begin{equation}
\begin{aligned}
 \bra \Tr_R U \ket=\Bigl\bra \Tr_R e^{\rt{\frac{\la}{2N}}M}\Bigr\ket_{\text{mm}},
\end{aligned} 
\label{eq:SYMvsMM}
\end{equation}
where $U$ is the 1/2 BPS Maldacena-Wilson loop \cite{Maldacena:1998im}
\begin{equation}
\begin{aligned}
 U=\text{P}\exp\left[\oint_C ds\Bigl(A_\mu \dot{x}^\mu +\ri\Phi  |\dot{x}|\Bigr) \right],
\end{aligned} 
\end{equation}
with $\Phi$ being one of the six adjoint scalar fields in $\mathcal{N}=4$ SYM. 
The contour $C$ is a circle.
In \eqref{eq:SYMvsMM}, the expectation value $\bra \cdots\ket$ on the left hand side
is taken in the $\mathcal{N}=4$ SYM while that on the right hand side
$\bra \cdots\ket_{\text{mm}}$ is defined by a hermitian matrix model
with a Gaussian potential
\begin{equation}
\begin{aligned}
 \bra \mathcal{O}\ket_{\text{mm}} =\frac{\int dM e^{- \Tr M^2}\mathcal{O}}{\int dM e^{- \Tr M^2}}.
\end{aligned} 
\label{eq:vev-mm}
\end{equation}
In the following we will mainly consider the expectation value of the 
operator of the form $\mathcal{O}=\det f(M)$ for some function $f(x)$
\begin{equation}
\begin{aligned}
 \Bigl\bra \det f(M)\Bigr\ket_{\text{mm}}=\Bigl\bra \prod_{i=1}^Nf(x_i)\Bigr\ket_{\text{mm}},
\end{aligned} 
\end{equation}
where $x_i~(i=1,\cdots,N)$ are the eigenvalues of the matrix $M$. 
As shown in \cite{Okuyama:2006jc}, the eigenvalue integral
of Gaussian matrix model can be written as a 
system of $N$ fermions in a harmonic potential
\begin{equation}
\begin{aligned}
\Bigl\bra \det f(M)\Bigr\ket_{\text{mm}}&= \frac{\int\prod_{i=1}^N dx_i e^{-x_i^2} f(x_i) \lap(x)^2}{\int\prod_{i=1}^N dx_i e^{-x_i^2} \lap(x)^2}
=\frac{1}{N!}\int \prod_{i=1}^N dx_i\Psi(x_1,\cdots,x_N)^2 \prod_{i=1}^Nf(x_i),
\end{aligned} 
\label{eq:fvev}
\end{equation}
where $\lap(x)$ denotes the Vandermonde determinant
and $\Psi(x_1,\cdots,x_N)$ is the Slater determinant of 
 harmonic oscillator wavefunctions
\begin{equation}
\begin{aligned}
 \Psi(x_1,\cdots,x_N)&=\det_{1\leq i,j\leq N}\Bigl(\psi_{i-1}(x_j)\Bigr),\\
\psi_n(x)&=\frac{1}{\rt{2^nn!\rt{\pi}}}H_n(x)e^{-\hf x^2},
\end{aligned} 
\end{equation}
with $H_n(x)$ being the Hermite polynomial.
The $N$-particle wavefunction $\Psi(x_1,\cdots,x_N)$ is also written as
\begin{equation}
\begin{aligned}
&\Psi(x_1,\cdots,x_N)= \bra x_1,\cdots,x_N|\Psi\ket,\\
&|\Psi\ket=|0\ket\wg |1\ket\wg\cdots\wg |N-1\ket,
\end{aligned} 
\label{eq:wg}
\end{equation}
where $|n\ket$ is the $n$-th excited state
of harmonic oscillator
\begin{equation}
\begin{aligned}
 |n\ket=\frac{(a^\dagger)^n}{\rt{n!}}|0\ket,\qquad [a,a^\dagger]=1.
\end{aligned} 
\end{equation}
In other words, $|\Psi\ket$ is the state of $N$ fermions
occupying the lowest $N$ levels of harmonic oscillator.
In terms of the state $|\Psi\ket$, the
expectation value in \eqref{eq:fvev}
can be also written as 
\begin{equation}
\begin{aligned}
\Bigl\bra \det f(M)\Bigr\ket_{\text{mm}}=\frac{1}{N!}\bra \Psi|\prod_{i=1}^N f\Bigl(\frac{a_i+a_i^\dagger}{\rt{2}}\Bigr)|\Psi\ket
\end{aligned} 
\label{eq:f-exp}
\end{equation}
where $[a_i,a_j^\dagger]=\cob_{i,j}$ and $a_i$ acts on the $i$-th factor of the state
$|\Psi\ket$ in \eqref{eq:wg}.
One can easily show that \eqref{eq:f-exp}
is further rewritten as a determinant of $N\times N$ matrix
\begin{equation}
\begin{aligned}
\Bigl\bra \det f(M)\Bigr\ket_{\text{mm}}=\det_{0\leq n,m\leq N-1}(f_{n,m}),
\end{aligned} 
\label{eq:fnm}
\end{equation}
where 
\begin{equation}
\begin{aligned}
 f_{n,m}=\Big\bra n\Big|f\Bigl(\frac{a+a^\dagger}{\rt{2}}\Bigr)\Big|m\Bigr\ket.
\end{aligned} 
\end{equation}

Now let us consider the correlator of Wilson loop $\Tr U^{k}$
with winding number $k$. The Wilson loop in the fundamental representation
corresponds to $k=1$.
The correlator of winding Wilson loops is also
 exactly written as
an expectation value in the Gaussian matrix model  \eqref{eq:vev-mm}
\begin{equation}
\begin{aligned}
 \Bigl\bra \prod_{i=1}^h \Tr U^{k_i}\Bigr\ket&=
\Bigl\bra \prod_{i=1}^h \Tr e^{k_i\rt{\frac{\la}{2N}}M}\Bigr\ket_{\text{mm}}.
\end{aligned} 
\label{eq:corr-mm}
\end{equation}
Note that our winding Wilson loops $\Tr U^{k_i}$ are located along the 
same circle
with the same radius. Namely, they are on top of each other in spacetime. 
In order to make use of the relation \eqref{eq:fnm},
we notice that the trace of matrix appears at the linear order
of the characteristic polynomial
\begin{equation}
\begin{aligned}
 \det\Big(1+y_ie^{k_i\rt{\frac{\la}{2N}}M}\Big)=1+y_i\Tr e^{k_i\rt{\frac{\la}{2N}}M}+\mathcal{O}(y_i^2).
\end{aligned} 
\end{equation}
Then we can rewrite \eqref{eq:corr-mm}
as a contour integral around $y_i=0$
\begin{equation}
\begin{aligned}
 \Bigl\bra \prod_{i=1}^h \Tr U^{k_i}\Bigr\ket=\oint\prod_{i=1}^h\frac{dy_i}{2\pi\ri y_i^2}G,
\end{aligned} 
\label{eq:corr-gen}
\end{equation}
where $G$ is given by
\begin{equation}
\begin{aligned}
 G&=\Biggl\bra \prod_{i=1}^h \det\Big(1+y_ie^{k_i\rt{\frac{\la}{2N}}M}\Big)\Biggr\ket_{\text{mm}}\\
&=\Biggl\bra \det\left(\sum_{m=0}^h\sum_{i_1<\cdots<i_m}y_{i_1}\cdots y_{i_m}
e^{\rt{\frac{\la}{2N}}(k_{i_1}+\cdots+ k_{i_m})M}\right)\Biggr\ket_{\text{mm}}.
\end{aligned} 
\end{equation}
Applying the relation \eqref{eq:fnm} to the above form of $G$, we find
\begin{equation}
\begin{aligned}
 G=\det\left(\sum_{m=0}^h\sum_{i_1<\cdots<i_m}y_{i_1}\cdots y_{i_m}
A(k_{i_1}+\cdots+ k_{i_m})\right),
\end{aligned} 
\end{equation}
where the matrix $A(k)$ is given by
\begin{equation}
\begin{aligned}
 A(k)_{n,m}=\bra n|e^{k\rt{\frac{\la}{4N}}(a+a^\dagger)}|m\ket
=\rt{\frac{n!}{m!}}e^{\frac{k^2\la}{8N}}\Bigl(\frac{k^2\la}{4N}\Bigr)^{\frac{m-n}{2}}
L_n^{m-n}\Bigl(-\frac{k^2\la}{4N}\Bigr),
\end{aligned} 
\end{equation}
and $L_n^{m-n}(z)$ denotes the associated Laguerre polynomial.
Note that $A(k)$ is a symmetric $N\times N$ matrix \footnote{It is emphasized in \cite{CanazasGaray:2018cpk} that the matrix $A(k)$ is closely related to the truncated harmonic oscillator,
which is widely used in quantum optics community. In our approach, the truncation arises
by taking the expectation value with respect to the ``Fermi sea'' state $|\Psi\ket$ 
\eqref{eq:f-exp}.}
\begin{equation}
\begin{aligned}
 A(k)_{n,m}=A(k)_{m,n},\quad (n,m=0,\cdots, N-1).
\end{aligned} 
\end{equation}

From \eqref{eq:corr-gen},
$G$ can be thought of as the generating function of the correlator of
winding Wilson loops. 
As we learned from quantum field theory textbooks,
the connected part of correlator can be extracted by taking the log of this
generating function $G$
\begin{equation}
\begin{aligned}
 \Bigl\bra \prod_{i=1}^h\Tr U^{k_i}\Bigr\ket_{\text{conn}}=\oint\prod_{i=1}^h\frac{dy_i}{2\pi\ri y_i^2}
\log G.
\end{aligned} 
\end{equation}
Finally, using the identity $\log\det X=\Tr\log X$,
we arrive at the exact finite $N$ expression of the connected correlator
of winding Wilson loops 
\begin{equation}
\begin{aligned}
 \Bigl\bra \prod_{i=1}^h\Tr U^{k_i}\Bigr\ket_{\text{conn}}=\oint\prod_{i=1}^h\frac{dy_i}{2\pi\ri y_i^2}
\Tr \log \left(\sum_{m=0}^h\sum_{i_1<\cdots<i_m}y_{i_1}\cdots y_{i_m}
A(k_{i_1}+\cdots+ k_{i_m})\right).
\end{aligned} 
\label{eq:exact-conn}
\end{equation}

One can easily see that the known expectation value of single Wilson loop \cite{Drukker:2000rr}
is reproduced from our exact result \eqref{eq:exact-conn}
\begin{equation}
\begin{aligned}
 \bra \Tr U^k\ket=\oint\frac{dy}{2\pi\ri y^2}\Tr\log\Big(1+yA(k)\Big)=\Tr A(k) .
\end{aligned} 
\end{equation}
The trace of $A(k)$ is evaluated as \cite{Drukker:2000rr}
\begin{equation}
\begin{aligned}
 \Tr A(k) 
=e^{\frac{k^2\la}{8N}}L_{N-1}^1\Bigl(-\frac{k^2\la}{4N}\Bigr).
\end{aligned} 
\end{equation}

In a similar manner, we can write down the connected higher point correlation functions of winding Wilson loops.
The two-point function is given by
\begin{equation}
\begin{aligned}
 \bra \Tr U^{k_1}\Tr U^{k_2}\ket_{\text{conn}}
=\Tr\Bigl[A(k_1+k_2)-A(k_1)A(k_2)\Bigr].
\end{aligned} 
\label{eq:2pt-finite}
\end{equation}
This reproduces the known result in \cite{Drukker:2000rr,Kawamoto:2008gp}. 
 
We can proceed to more higher point functions. For instance,
the three-point function is given by
\begin{equation}
\begin{aligned}
 &\bra \Tr U^{k_1}\Tr U^{k_2}\Tr U^{k_3}\ket_{\text{conn}}\\
=&\Tr\Bigl[A(k_1+k_2+k_3)+A(k_1)A(k_2)A(k_3)+A(k_1)A(k_3)A(k_2)\\
&\quad -
A(k_1)A(k_2+k_3)-A(k_2)A(k_1+k_3)-A(k_3)A(k_1+k_2)\Bigr],
\end{aligned} 
\label{eq:3pt-finite}
\end{equation}
and the four-point function is given by
\begin{equation}
\begin{aligned}
&\bra \Tr U^{k_1}\Tr U^{k_2}\Tr U^{k_3}\Tr U^{k_4}\ket_{\text{conn}}\\
=&\Tr\Bigl[
A(k_1+k_2+k_3+k_4)\\
&+A(k_1+k_2)\{A(k_3),A(k_4)\}+A(k_1+k_3)\{A(k_2),A(k_4)\}\\
&+A(k_1+k_4)\{A(k_2),A(k_3)\}
+A(k_2+k_3)\{A(k_1),A(k_4)\}\\
&+A(k_2+k_4)\{A(k_1),A(k_3)\}+A(k_3+k_4)\{A(k_1),A(k_2)\}\\
&-A(k_1+k_2)A(k_3+k_4)-A(k_1+k_3)A(k_2+k_4)-A(k_1+k_4)A(k_2+k_3)\\
&-A(k_1)A(k_2+k_3+k_4)-A(k_2)A(k_1+k_3+k_4)\\
&-A(k_3)A(k_1+k_2+k_4)-A(k_4)A(k_1+k_2+k_3)\\
&-A(k_1)A(k_2)A(k_3)A(k_4)-A(k_1)A(k_3)A(k_2)A(k_4)\\
&-A(k_2)A(k_1)A(k_3)A(k_4)-A(k_2)A(k_3)A(k_1)A(k_4)\\
&-A(k_3)A(k_1)A(k_2)A(k_4)-A(k_3)A(k_2)A(k_1)A(k_4)
\Bigr]
\end{aligned} 
\label{eq:4pt-finite}
\end{equation}
where $\{A,B\}$ denotes the anti-commutator: $\{A,B\}=AB+BA$.

\section{$1/N$ expansion from topological recursion \label{sec:top}}
As demonstrated in \cite{Okuyama:2017feo},
the $1/N$ correction of 1/2 BPS Wilson loops can be computed systematically 
using the topological recursion of Gaussian matrix model.
Let us consider the $1/N$ expansion of connected correlator
\begin{equation}
\begin{aligned}
 \Bigl\bra \prod_{i=1}^h \Tr U^{k_i}\Bigr\ket_{\text{conn}}
=\sum_{g=0}^\infty N^{2-2g-h}\mathcal{C}_{g,h}(k_1,\cdots,k_h).
\end{aligned} 
\label{eq:genus-conn}
\end{equation}
The genus-$g$ contribution to the $h$-point connected correlator 
$\mathcal{C}_{g,h}$ is written as
\begin{equation}
\begin{aligned}
 \mathcal{C}_{g,h}(k_1,\cdots,k_h)=\int \prod_{i=1}^h du_i \rho_{g,h}(u_1,\cdots,u_h)
\prod_{i=1}^h e^{\rt{\la}k_iu_i},
\end{aligned} 
\label{eq:C-rho-int}
\end{equation}
where the multi-point density $\rho_{g,h}(u_1,\cdots,u_h)$
is obtained from the discontinuity of the connected correlator
$W_{g,h}(x_i)$ of the resolvents in Gaussian matrix model
\begin{equation}
\begin{aligned}
 \left\bra\prod_{i=1}^h\Tr \frac{1}{x_i-M}\right\ket_{\text{conn}}=
\sum_{g=0}^\infty N^{2-2g-h}W_{g,h}(x_1,\cdots,x_h).
\end{aligned} 
\end{equation} 
Here we have rescaled the matrix $M\to \rt{2N}M$ so that the 
measure of Gaussian matrix model becomes
$\int dM e^{-2N\Tr M^2}$. In this normalization, the eigenvalues are 
distributed along the cut $[-1,1]$ and the eigenvalue density is 
given by the Wigner semi-circle distribution
\begin{equation}
\begin{aligned}
 \rho_{0,1}(u)=\frac{2}{\pi}\rt{1-u^2}.
\end{aligned} 
\end{equation}
More generally, the multi-point density $\rho_{g,h}$ in \eqref{eq:C-rho-int}
is obtained from $W_{g,h}$ by taking the discontinuity across the cut $u_i\in[-1,1]$
for all variables $(x_1,\cdots,x_h)$
\begin{equation}
\begin{aligned}
\rho_{g,h}(u_1,\cdots,u_h)=\left(\prod_{i=1}^h \text{Disc}_i\right) W_{g,h} ,
\end{aligned} 
\label{eq:disc}
\end{equation}
where $\text{Disc}_i$ denotes the discontinuity of the $i$-th variable $x_i$
\begin{equation}
\begin{aligned}
 \text{Disc}_iW_{g,h}=\frac{W_{g,h}(x_i=u_i+\ri 0)-W_{g,h}(x_i=u_i-\ri 0)}{2\pi\ri}.
\end{aligned} 
\end{equation}
As shown in \cite{Eynard:2004mh}\footnote{See also 
 \cite{Eynard:2008we} for a review of topological recursion.}, $W_{g,h}$ is determined recursively by the 
following relation
\begin{align}
 \begin{aligned}
  4x_1W_{g,h}(x_1,\cdots,x_h)=&W_{g-1,h+1}(x_1,x_1,x_2,\cdots,x_h)+4\cob_{g,0}\cob_{h,1}\\
&+\sum_{I_1\sqcup I_2=\{2,\cdots,h\}}\sum_{g'=0}^gW_{g',1+|I_1|}(x_1,x_{I_1})
W_{g-g',1+|I_2|}(x_1,x_{I_2})
\\
&+\sum_{j=2}^h \frac{\del}{\del x_j}\frac{W_{g,h-1}(x_1,\cdots,\h{x}_j,\cdots,x_h)
-W_{g,h-1}(x_2,\cdots,x_h)}{x_1-x_j}.
 \end{aligned}
\end{align}
Using this recursion relation, one can easily compute $W_{g,h}$
starting from $W_{0,1}$
\begin{equation}
\begin{aligned}
 W_{0,1}(x)=\int_{-1}^1 du \rho_{0,1}(u)\frac{1}{x-u}=2x-2\rt{1-x^2}.
\end{aligned} 
\end{equation}

\subsection{Two-point function}
Let us compute the genus-$g$
contribution of the two-point function $\mathcal{C}_{g,2}$.
The genus-zero part $\mathcal{C}_{0,2}$ is exceptional:
it is not given by the general expression \eqref{eq:C-rho-int}
but  $\mathcal{C}_{0,2}$ is written as \cite{Haagerup}
\begin{equation}
\begin{aligned}
 \mathcal{C}_{0,2}(k_1,k_2)=
\frac{1}{4\pi^2}\int_{-1}^1du\int_{-1}^1dv
\frac{1-uv}{\rt{(1-u^2)(1-v^2)}}\frac{(e^{k_1\rt{\la}u}-e^{k_1\rt{\la}v})
(e^{k_2\rt{\la}u}-e^{k_2\rt{\la}v})}{(u-v)^2}.
\end{aligned} 
\end{equation}
As shown in \cite{Akemann:2001st,Giombi:2009ms}, this integral can be evaluated in a closed form
in terms of the modified Bessel function of the first kind $I_\nu(x)$
\begin{equation}
\begin{aligned}
 \mathcal{C}_{0,2}(k_1,k_2)=\frac{\rt{\la}k_1k_2}{2(k_1+k_2)}\Bigl[I_0(k_1\rt{\la})I_1(k_2\rt{\la})
+I_1(k_1\rt{\la})I_0(k_2\rt{\la})\Bigr].
\end{aligned} 
\label{eq:C02}
\end{equation} 

Next consider the genus-one part $\mathcal{C}_{1,2}$.
From the topological recursion, $W_{1,2}$ is found to be
\begin{equation}
\begin{aligned}
 W_{1,2}=\frac{5(1+x_1x_2)}{64}\Bigl(\frac{1}{r_1^3r_2^7}+\frac{1}{r_1^7r_2^3}\Bigr)
+\frac{3(1+x_1x_2)}{64r_1^5r_2^5}+\frac{1}{16}\Bigl(\frac{1}{r_1^3r_2^5}+\frac{1}{r_1^5r_2^3}\Bigr),
\end{aligned} 
\label{eq:W12-ri}
\end{equation}
where we introduced the notation $r_i$ by
\begin{equation}
\begin{aligned}
 r_i=\rt{x_i^2-1}.
\end{aligned} 
\end{equation}
As discussed in \cite{Okuyama:2017feo}, in order to 
compute $\mathcal{C}_{g,h}$ in \eqref{eq:C-rho-int}
we need to  rewrite \eqref{eq:W12-ri} using the formula
\begin{equation}
\begin{aligned}
 \frac{1}{r_i^{2n+1}}&=\frac{(-1)^n}{(2n-1)!!}\del_i^n t_n(x_i),\\
 \frac{x_i}{r_i^{2n+1}}&=\frac{(-1)^n}{(2n-1)!!}\del_i^n t_{n-1}(x_i),
\end{aligned} 
\label{eq:r-t}
\end{equation}
where $\del_i=\frac{\del}{\del x_i}$ and
$t_n(x_i)$ is related to the Chebychev polynomial of the first kind $T_n(x_i)$
by
\begin{equation}
\begin{aligned}
 t_n(x_i)=\frac{T_n(x_i)}{r_i}.
\end{aligned} 
\end{equation}
Then $W_{1,2}$ in \eqref{eq:W12-ri} is written in terms of $t_n$ as
\begin{equation}
\begin{aligned}
 W_{1,2}&=\frac{1}{192}\Bigl[\del_1\del_2^3(t_1t_3+t_0t_2)+(x_1\leftrightarrow x_2)\Bigr]
+\frac{1}{192} \del_1^2\del_2^2(t_2t_2+t_1t_1)\\
&\quad -\frac{1}{48}\Bigl[\del_1\del_2^2t_1t_2+(x_1\leftrightarrow x_2)\Bigr],
\end{aligned} 
\label{eq:W12}
\end{equation}
where the product $t_nt_m$ should be understood as $t_n(x_1)t_m(x_2)$.

Now, $\mathcal{C}_{1,2}$ is obtained from $W_{1,2}$ in \eqref{eq:W12}
using the prescription \eqref{eq:C-rho-int} and \eqref{eq:disc}.
The $u$-integral in \eqref{eq:C-rho-int} 
is easily evaluated by the formula
\begin{equation}
\begin{aligned}
 \int_{-1}^1 du \frac{T_n(u)}{\pi\rt{1-u^2}} e^{k\rt{\la}u}=I_n(k\rt{\la}).
\end{aligned} 
\end{equation}
After performing the integration by parts, 
we find that $\mathcal{C}_{1,2}$ is given by
\begin{equation}
\begin{aligned}
 \mathcal{C}_{1,2}(k_1,k_2)=
\frac{\la^2}{192}\Bigl[
&k_1k_2^3I_1(k_1\rt{\la})I_3(k_2\rt{\la})+k_1k_2^3I_0(k_1\rt{\la})I_2(k_2\rt{\la})\\
+&
k_1^3k_2I_3(k_1\rt{\la})I_1(k_2\rt{\la})+k_1^3k_2I_2(k_1\rt{\la})I_0(k_2\rt{\la})\\
+&k_1^2k_2^2I_2(k_1\rt{\la})I_2(k_2\rt{\la})+k_1^2k_2^2I_1(k_1\rt{\la})I_1(k_2\rt{\la})\Bigr]\\
+\frac{\la^{3/2}}{48}\Bigl[&
k_1k_2^2I_1(k_1\rt{\la})I_2(k_2\rt{\la})
+k_1^2k_2I_2(k_1\rt{\la})I_1(k_2\rt{\la})\Bigr].
\end{aligned} 
\label{eq:C12}
\end{equation}

In a similar manner, we can in principle compute the $1/N$ genus expansion up to any desired order.
For instance, the genus-two correction $W_{2,2}$ is given by
\begin{equation}
\begin{aligned}
 W_{2,2}&=\frac{11}{30720}\del_1^3\del_2^3(5t_3t_3+2t_2t_2)\\
+&\Bigl[-\frac{29}{184320}\del_1^3\del_2^4(t_3t_4+t_2t_3)
-\frac{1}{12288}\del_1^2\del_2^5(t_2t_5+t_1t_4)
+\frac{7}{46080}\del_1^2\del_2^4(8t_2t_4+3t_1t_3)\\
&-\frac{13}{3840}\del_1^2\del_2^3t_2t_3
-\frac{1}{640}\del_1\del_2^4t_1t_4
+\frac{1}{92160}\del_1\del_2^5(43t_1t_5+18t_0t_4)
-\frac{1}{36864}\del_1\del_2^6(t_1t_6+t_0t_5)\\
&\qquad +(x_1\leftrightarrow x_2)\Bigr],
\end{aligned} 
\end{equation}
and this can be translated to the genus-two correction
 of the two-point function $\mathcal{C}_{2,2}(k_1,k_2)$
using 
the prescription \eqref{eq:C-rho-int} and \eqref{eq:disc}. For simplicity, 
here we write down  
the expression for $k_1=k_2=1$
\begin{equation}
\begin{aligned}
 &\quad \mathcal{C}_{2,2}(1,1)=\\
&\frac{29 \la^{7/2} \left(I_2(\rt{\la})
   I_3(\rt{\la})+I_4(\rt{\la})
   I_3(\rt{\la})\right)}{92160}+\frac{\la^{7/2} \left(I_1(\rt{\la})
   I_4(\rt{\la})+I_2(\rt{\la})
   I_5(\rt{\la})\right)}{6144}\\
+&\frac
   {\la^{7/2} \left(I_0(\rt{\la})
   I_5(\rt{\la})+I_1(\rt{\la})
   I_6(\rt{\la})\right)}{18432}\\
+&\frac{13 \la^{5/2} I_2(\rt{\la})
   I_3(\rt{\la})}{1920}
+\frac{1}{320
   } \la^{5/2} I_1(\rt{\la})
   I_4(\rt{\la})
+\frac{11 \la^3
   \left(2 I_2(\rt{\la}){}^2+5
   I_3(\rt{\la}){}^2\right)}{30720}\\
+&
   \frac{7 \la^3 \left(3 I_1(\rt{\la})
   I_3(\rt{\la})+8
   I_2(\rt{\la})
   I_4(\rt{\la})\right)}{23040}
+\frac{\la^3 \left(18 I_0(\rt{\la})
   I_4(\rt{\la})+43
   I_1(\rt{\la})
   I_5(\rt{\la})\right)}{46080}.
\end{aligned} 
\label{eq:C22}
\end{equation}

\subsection{Three-point and four-point functions}
We can compute the three-point and four-point functions in a similar way.
For simplicity, we write down the genus-zero part when all winding numbers are unity: $k_i=1~(i=1,\cdots,h)$.
We find that the genus-zero part of the three-point function is given by\footnote{We can write down the 
three point function $\mathcal{C}_{0,3}(k_1,k_2,k_3)$ for general winding numbers  using the 
explicit form of $W_{0,3}$ in \cite{Okuyama:2017feo}, which agrees with the known result of 
$\mathcal{C}_{0,3}(k_1,k_2,k_3)$ in \cite{Akemann:2001st}.}
\begin{equation}
\begin{aligned}
 \mathcal{C}_{0,3}(1,1,1)=\frac{\la^{\frac{3}{2}}}{8}
\Bigl[3I_0(\rt{\la})^2I_1(\rt{\la})+
I_1(\rt{\la})^3\Bigr],
\end{aligned} 
\label{eq:C03}
\end{equation}
and the genus-zero part of the four-point function is given by
\begin{equation}
\begin{aligned}
 &\mathcal{C}_{0,4}(1,\cdots,1)=\frac{3\la^2}{16}\Biggl[4I_0(\rt{\la})^2I_1(\rt{\la})^2+I_1(\rt{\la})^4\Biggr]\\
+&\frac{\la^{\frac{5}{2}}}{8}\Biggl[3I_0(\rt{\la})^2I_1(\rt{\la})I_2(\rt{\la})+I_1(\rt{\la})^3
I_2(\rt{\la})
+3I_0(\rt{\la})I_1(\rt{\la})^3+I_0(\rt{\la})^3I_1(\rt{\la})\Biggr].
\end{aligned}
\label{eq:C04} 
\end{equation}

\section{Comparison between the exact result and the $1/N$ expansion \label{sec:comparison}}
In this section, we compare the exact finite $N$ result in section 
\ref{sec:exact} and the 
$1/N$ expansion obtained from the topological recursion in section \ref{sec:top}.
\subsection{Small $\la$ regime}
Lets us consider the two-point function with $k_1=k_2=1$.
The exact result \eqref{eq:2pt-finite} in this case reads
\begin{equation}
\begin{aligned}
 \bra (\Tr U)^2\ket_{\text{conn}}=\Tr[A(2)-A(1)^2].
\end{aligned} 
\label{eq:2pt-11}
\end{equation}
In this subsection we consider the small $\la$ expansion of this exact result.
It turns out that the each term in the small $\la$ expansion
receives only a finite number of $1/N$ corrections
\begin{equation}
\begin{aligned}
 \bra (\Tr U)^2\ket_{\text{conn}}=
&\frac{1}{4}\la+
\frac{3}{32}\la^2+\left(\frac{5}{384}+\frac{1}{192 N^2}\right)\la^3
+\left(\frac{35}{36864}+\frac{55}{36864 N^2}\right)\la^4\\
&+\left(\frac{7}{163840}+\frac{49}{294912N^2}+\frac{1}{23040 N^4}\right)\la^5\\
&+\left(\frac{77}{58982400}+\frac{119}{11796480
   N^2}+\frac{49}{4915200
   N^4}\right)\la^6+\mathcal{O}(\la^7).
\end{aligned} 
\label{eq:2pt-small-la}
\end{equation}
The coefficient of $\mathcal{O}(\la^n)$ term
can be easily obtained from the 
small $\la$ expansion of the exact result \eqref{eq:2pt-11}
at first few $N$'s $(N=1,2,\cdots,[(n+1)/2])$.
This expansion \eqref{eq:2pt-small-la}
can be reorganized into the form of the genus expansion \eqref{eq:genus-conn}
\begin{equation}
\begin{aligned}
 \bra (\Tr U)^2\ket_{\text{conn}}=\mathcal{C}_{0,2}+\frac{1}{N^2}
\mathcal{C}_{1,2}+\frac{1}{N^4}
\mathcal{C}_{2,2}+\mathcal{O}(N^{-6})
\end{aligned} 
\end{equation}
with
\begin{equation}
\begin{aligned}
 \mathcal{C}_{0,2}&=\frac{1}{4}\la
+\frac{3}{32}\la^2+\frac{5}{384}\la^3+\frac{35}{36864}\la^4+\frac{7}{163840}\la^5
+\frac{77}{58982400}\la^6+\mathcal{O}(\la^7),\\
\mathcal{C}_{1,2}&=\frac{1}{192}\la^3+\frac{55}{36864}\la^4+
\frac{49}{294912}\la^5+
\frac{119}{11796480}\la^6+\mathcal{O}(\la^7),\\
\mathcal{C}_{2,2}&=\frac{1}{23040}\la^5+\frac{49}{4915200}\la^6+
\mathcal{O}(\la^7).
\end{aligned} 
\end{equation}
As expected, this agrees with 
the small $\la$ expansion of $\mathcal{C}_{0,2}$ in \eqref{eq:C02},
$\mathcal{C}_{1,2}$ in \eqref{eq:C12}, 
and $\mathcal{C}_{2,2}$ in \eqref{eq:C22} obtained from 
the topological recursion. 

We have performed a similar test for the 
three-point function and the four-point function at genus-zero and find
perfect agreement between the exact finite $N$ result \eqref{eq:3pt-finite}, \eqref{eq:4pt-finite} and the analytic result 
\eqref{eq:C03}, \eqref{eq:C04} obtained from the topological recursion.

\subsection{Finite $\la$ regime}
In the finite $\la$ regime, 
we can test numerically the agreement 
between the finite $N$ result and the $1/N$ expansion 
obtained from the topological recursion.

We can extract the genus-$g$ contribution
$\mathcal{C}_{g,h}$ from the exact
finite $N$ result in \eqref{eq:exact-conn} assuming that $\mathcal{C}_{g',h}$ with
$g'\leq g-1$ are already known
\begin{equation}
\begin{aligned}
 \mathcal{C}_{g,h}\approx N^{-2+2g+h}\left[\Bigr\bra \prod_{i=1}^h \Tr U^{k_i}\Bigr\ket_{\text{conn}}
-\sum_{g'=0}^{g-1}N^{2-2g'-h}\mathcal{C}_{g',h}\right],\quad(N\gg1).
\end{aligned} 
\label{eq:C-exact}
\end{equation}
Then we can compare this with 
the analytic form of $\mathcal{C}_{g,h}$ obtained from the topological recursion.
Once we have checked that $\mathcal{C}_{g,h}$ in \eqref{eq:C-exact}
agrees with the analytic result, we can
proceed to check the next order $\mathcal{C}_{g+1,h}$ by plugging the analytic result of 
$\mathcal{C}_{g,h}$ into the right hand side of \eqref{eq:C-exact}.

\begin{figure}[htb]
\centering
\subcaptionbox{$\mathcal{C}_{0,2}$\label{sfig:C02}}{\includegraphics[width=4.8cm]{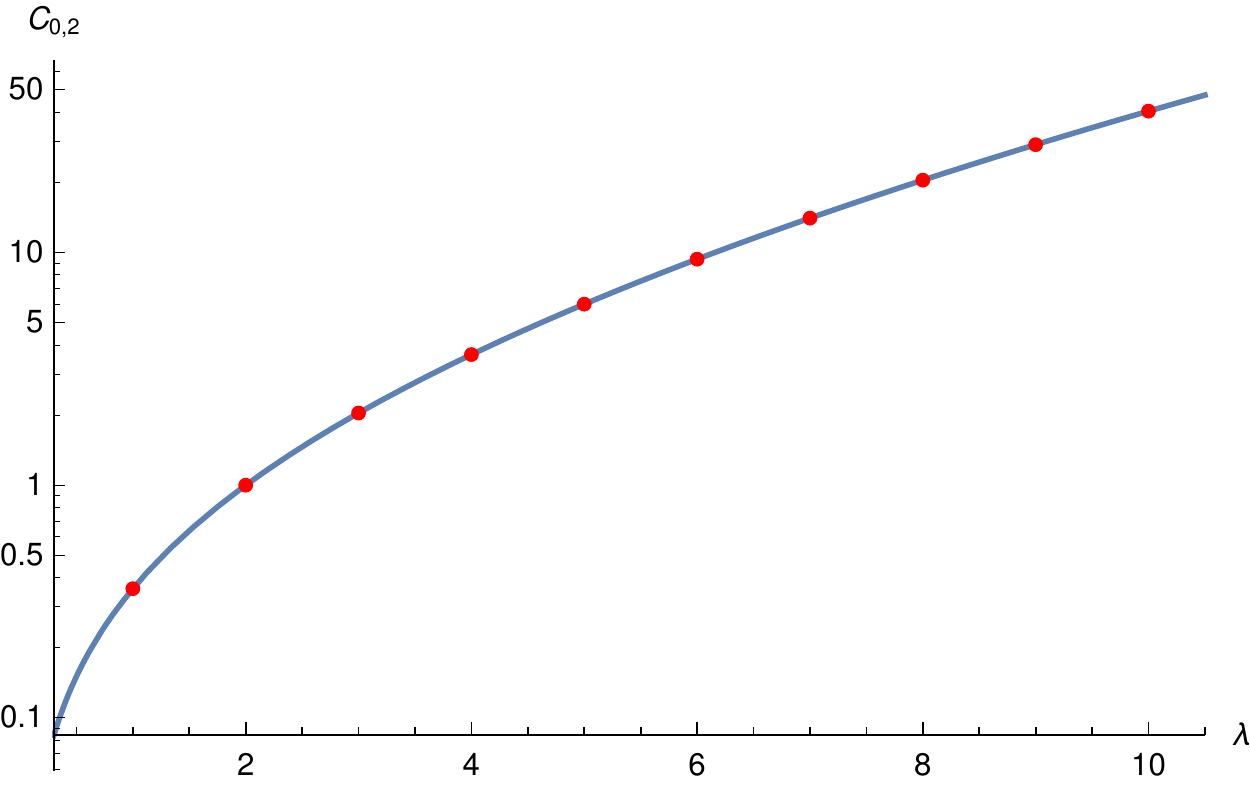}}
\hskip2mm
\subcaptionbox{$\mathcal{C}_{1,2}$\label{sfig:C12}}{\includegraphics[width=4.8cm]{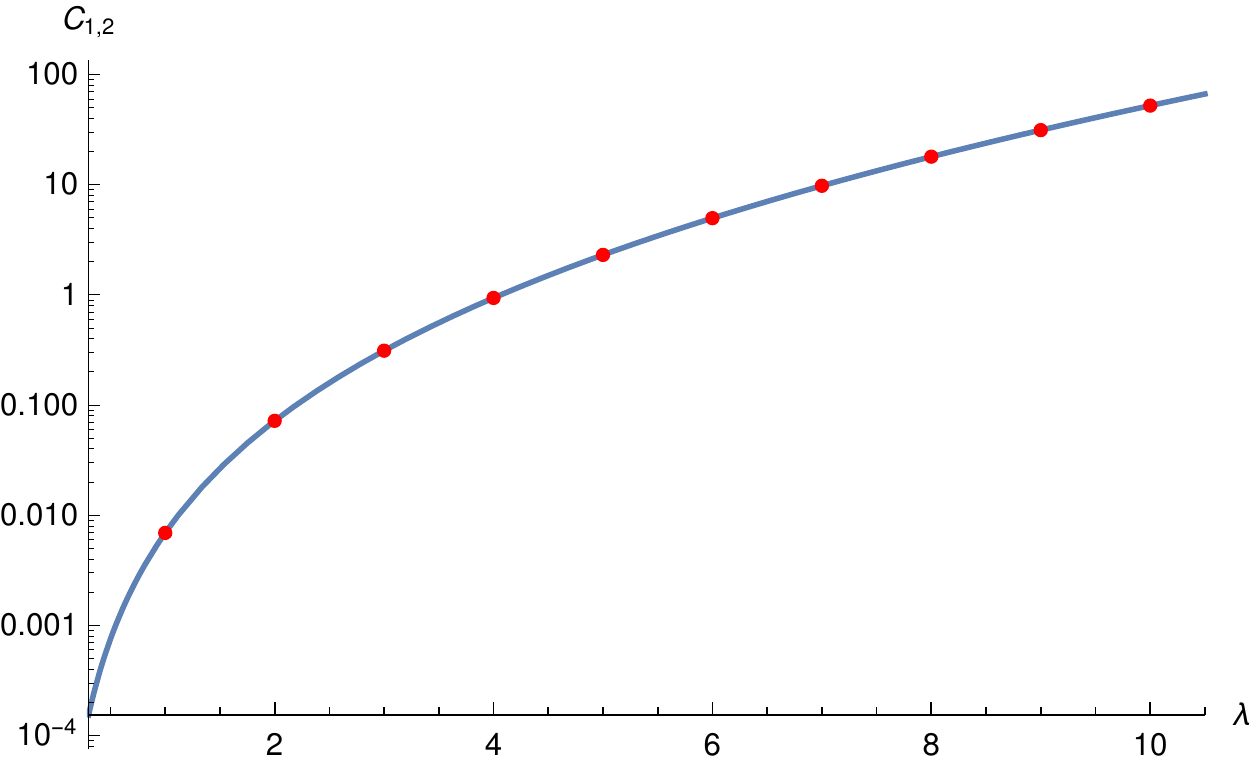}}
\hskip2mm
\subcaptionbox{$\mathcal{C}_{2,2}$\label{sfig:C22}}{\includegraphics[width=4.8cm]{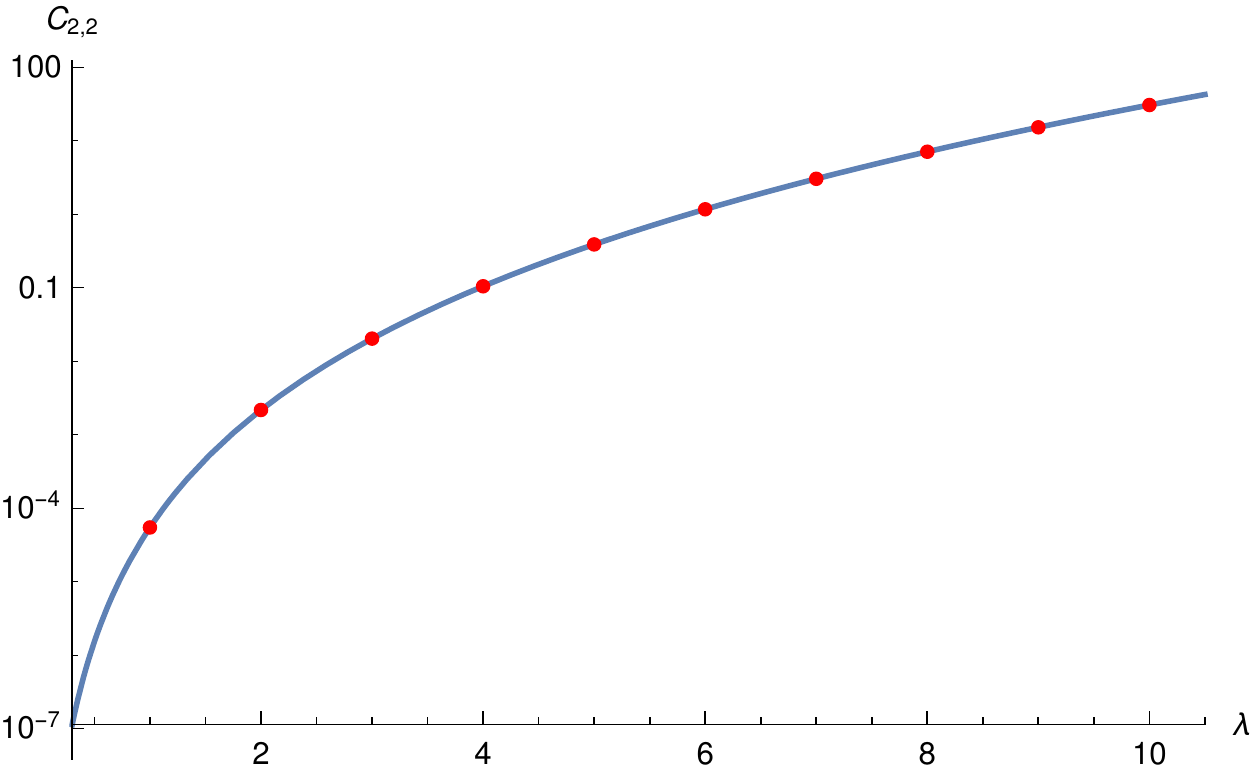}}
  \caption{
Plot 
of two-point function $\mathcal{C}_{g,2}$ as a function of $\la$ for \subref{sfig:C02} $g=0$,
\subref{sfig:C12} $g=1$, and \subref{sfig:C22} $g=2$.
The red dots are the exact values of the right hand side of
\eqref{eq:C-exact} at $N=200$. The blue solid curves are 
the analytic result of $\mathcal{C}_{g,2}$ obtained from the topological recursion.
}
  \label{fig:2pt}
\end{figure}

In Figure~\ref{fig:2pt}, we show the result of  
this comparison for the two-point function with $k_1=k_2=1$.
As we can see from this figure, the $1/N$ corrections correctly match
between the exact finite $N$ result \eqref{eq:2pt-11} and the analytic result 
in \eqref{eq:C02},
\eqref{eq:C12}, 
and \eqref{eq:C22} obtained
from the topological recursion.

\begin{figure}[htb]
\centering
\subcaptionbox{$\mathcal{C}_{0,3}$\label{sfig:C03}}{\includegraphics[width=5cm]{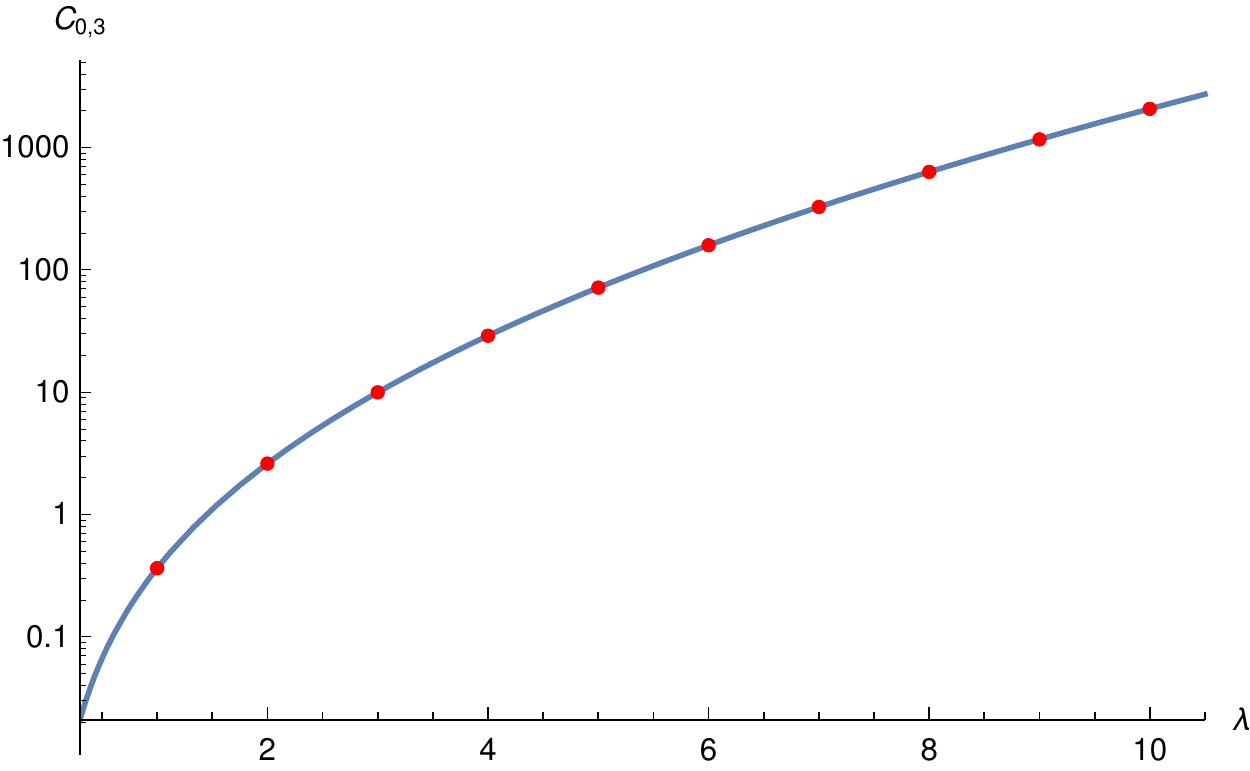}}
\hskip15mm
\subcaptionbox{$\mathcal{C}_{0,4}$\label{sfig:C04}}{\includegraphics[width=5cm]{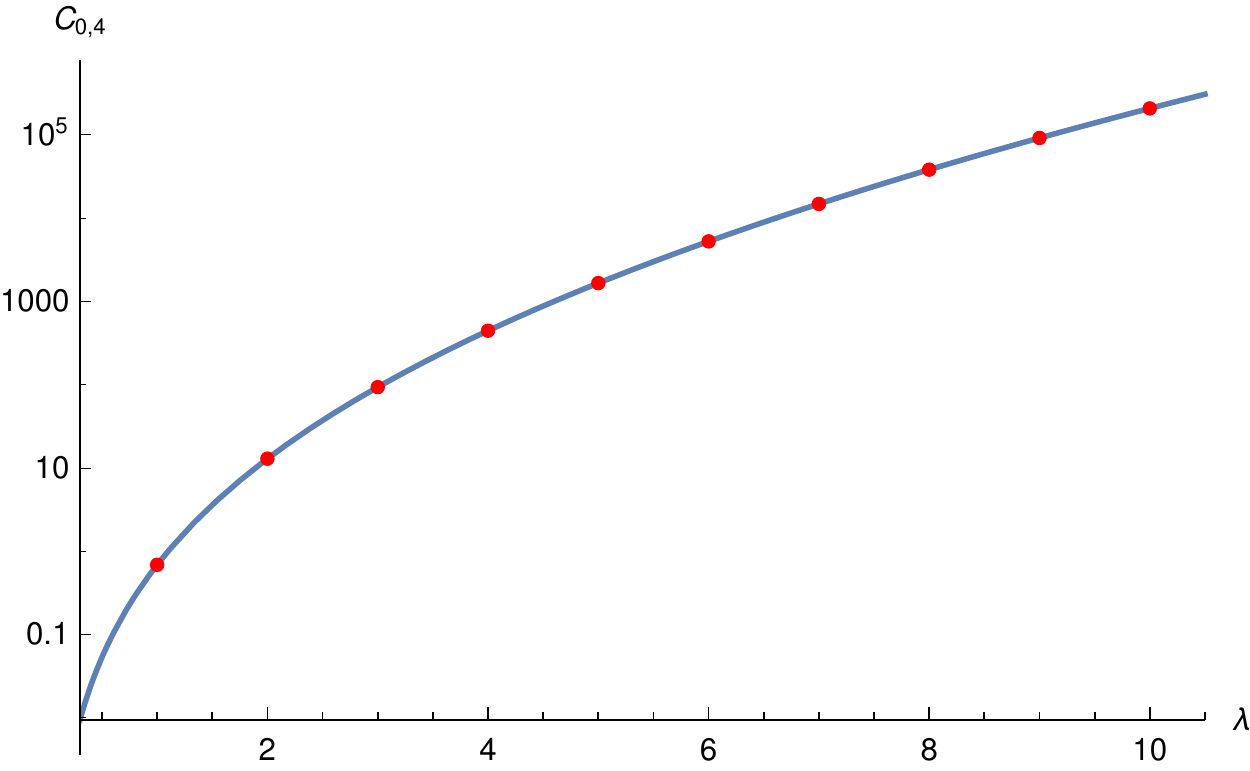}}
  \caption{ \subref{sfig:C03} and \subref{sfig:C04} are the
plots 
of genus-zero three-point function $\mathcal{C}_{0,3}$ 
and  four-point function $\mathcal{C}_{0,4}$, respectively.
The red dots are the exact values of the right hand side of
\eqref{eq:C-exact} at $N=200$, while the blue solid curves are 
the analytic result of $\mathcal{C}_{0,3}$ and 
$\mathcal{C}_{0,4}$ obtained from the topological recursion.
}
  \label{fig:34pt}
\end{figure}

In Figure~\ref{fig:34pt}, we show the similar plot for the
genus-zero part of the three-point function $\mathcal{C}_{0,3}$ 
and the four-point function $\mathcal{C}_{0,4}$ with all winding numbers 
set to $k_i=1$.
Again, we find a perfect agreement between the exact finite $N$ result \eqref{eq:3pt-finite}, \eqref{eq:4pt-finite} and the
analytic result \eqref{eq:C03}, \eqref{eq:C04} of the topological recursion.

\section{Comment on large winding number \label{sec:large}}
When the winding number $k$ becomes of the order of $N$,
we can take a scaling limit where the following combination of $\ka$ 
is held fixed \cite{Drukker:2005kx}
\begin{equation}
\begin{aligned}
k,N\to \infty\quad\text{with}\quad \ka=\frac{k\rt{\la}}{4N}~~\text{fixed}.
\end{aligned} 
\label{eq:ka-lim}
\end{equation}
In this limit, the holographically dual object of winding Wilson loop is not
a fundamental string but a D3-brane \cite{Drukker:2005kx}.
In this limit, the expectation value scales as
\begin{equation}
\begin{aligned}
 \bra \Tr U^k\ket&=\Tr A(k) \approx  e^{NF(\ka)},
\end{aligned} 
\label{eq:D3}
\end{equation}
where $F(\ka)$ is identified with the on-shell action of D3-brane
\begin{equation}
\begin{aligned}
F(\ka)=2\ka\rt{\ka^2+1}+2\,\text{arcsinh}(\ka) .
\end{aligned} 
\end{equation}
For the connected two-point function of large winding loops,
we observed numerically that the first term in \eqref{eq:2pt-finite}
is dominant if
the two loops are \textit{parallel}, i.e. $k_1$ and $k_2$ are both positive.
For instance, when $k_1=k_2=k>0$, we find
\begin{equation}
\begin{aligned}
 \bra \Tr U^{k}\Tr U^{k}\ket_{\text{conn}}\approx \Tr A(2k)\approx
e^{NF(2\ka)}.
\end{aligned} 
\label{eq:para}
\end{equation} 
On the other hand, the \textit{anti-parallel} loops with opposite sign of windings
are considered in \cite{Giombi:2009ms}.
We observed numerically that
the second term in \eqref{eq:2pt-finite} is dominant in this case
\begin{equation}
\begin{aligned}
 \bra \Tr U^{k}\Tr U^{-k} \ket_{\text{conn}}&\approx -\Tr A(k)A(-k)\approx -e^{2NF(\ka)}.
\end{aligned} 
\label{eq:anti-para}
\end{equation}
In the bulk D3-brane picture, $k$ corresponds to the electric flux on the worldvolume of D3-brane \cite{Drukker:2005kx}.
Comparing \eqref{eq:D3} and \eqref{eq:para}, the parallel case \eqref{eq:para} seems to
correspond to a D3-brane with twice as much electric flux 
as the one for the single  loop $\bra \Tr U^k\ket$ in \eqref{eq:D3}.
On the other hand, the anti-parallel case \eqref{eq:anti-para}
seems to correspond to two D3-branes.
It would be interesting to understand the bulk D-brane picture 
of parallel and anti-parallel cases more clearly.

\section{Generating function of symmetric representations \label{sec:sym}}
Using our formalism \eqref{eq:fnm}, we can easily write down the
generating function of 1/2 BPS Wilson loops in
the $k$-th symmetric representation $S_k$ 
\begin{equation}
\begin{aligned}
 \sum_{k=0}^\infty z^k \bra\Tr_{S_k}U\ket=
\left\bra \det\Biggl(\frac{1}{1-ze^{\rt{\frac{\la}{2N}}M}}\Biggr)\right\ket_{\text{mm}}
=\left\bra \det\Biggl(1+\sum_{k=1}^\infty z^k e^{k\rt{\frac{\la}{2N}}M}\Biggr)
\right\ket_{\text{mm}}.
\end{aligned} 
\end{equation}
From \eqref{eq:fnm}, this is written as a determinant of $N\times N$
matrix
\begin{equation}
\begin{aligned}
 \sum_{k=0}^\infty z^k \bra\Tr_{S_k}U\ket=\det\left(1+\sum_{k=1}^\infty z^k A(k)\right).
\end{aligned} 
\label{eq:gen-sym}
\end{equation}
From this generating function, we can in principle study the
behavior of Wilson loop in the symmetric representation at finite $N$.
In particular, it would be interesting to study the 
$1/N$ correction along the lines of \cite{CanazasGaray:2018cpk}.

In the large winding number limit \eqref{eq:ka-lim}, 
it turns out that the dominant contribution 
to $\bra\Tr_{S_k}U\ket$
comes from the single trace Wilson loop with winding number $k$
\begin{equation}
\begin{aligned}
 \bra\Tr_{S_k}U\ket\approx \Tr A(k)\approx e^{NF(\ka)}.
\end{aligned} 
\end{equation}
The difference between  $\bra\Tr_{S_k}U\ket$ and $\bra \Tr U^k\ket=\Tr A(k)$
is exponentially suppressed in the limit \eqref{eq:ka-lim}.

Now let us compare the generating function of symmetric Wilson loops
\eqref{eq:gen-sym} and that of the anti-symmetric representations $A_k$
found in \cite{Fiol:2013hna}
\begin{equation}
\begin{aligned}
 \sum_{k=0}^\infty z^k \bra \Tr_{A_k}U\ket=
\Bigl\bra \det\Bigl(1+ze^{\rt{\frac{\la}{2N}}M}\Bigr)\Bigr\ket_{\text{mm}}=
\det\Big(1+zA(1)\Big).
\end{aligned} 
\label{eq:gen-asym}
\end{equation}
As we will see below, we find an interesting 
relation between  the log of the generating functions
\eqref{eq:gen-sym} and 
\eqref{eq:gen-asym}
\begin{equation}
\begin{aligned}
 J_S\Bigl(z,\la,\frac{1}{N}\Bigr)&=\frac{1}{N}\Tr\log \Biggl(1+\sum_{k=1}^\infty z^k A(k)\Biggr),
\\
J_A\Bigl(z,\la,\frac{1}{N}\Bigr)&=\frac{1}{N}\Tr\log \Bigl(1+zA(1)\Bigr).
\end{aligned} 
\end{equation}
As we have seen in section \ref{sec:comparison},
the coefficient of the small $\la$ expansion receives
only a finite number of $1/N$ corrections and hence
it can be determined from the small $\la$ expansion of
exact finite $N$ result for the first few $N$'s.
From the exact finite $N$ result in \eqref{eq:gen-asym},
the $1/N$ expansion of $J_A$ was obtained in \cite{Okuyama:2017feo}
in the small $\la$ regime
\begin{equation}
\begin{aligned}
 J_A\Bigl(z,\la,\frac{1}{N}\Bigr)&=\sum_{n=0}^\infty N^{-n}J^{(n)}_A(z,\la)
\end{aligned} 
\end{equation}
where the lower order terms are given by 
\begin{equation}
\begin{aligned}
J^{(0)}_A(z,\la)&=
\log(1+z)+\frac{z}{8(1+z)^2}\la+\frac{z(1-4z+z^2)}{192(1+z)^4}\la^2
+\frac{z \left(z^4-26 z^3+66 z^2-26 z+1\right)}{9216
   (1+z)^6}\la^3+\mathcal{O}(\la^4), \\
J^{(1)}_A(z,\la)&=
\frac{z^2}{8 (1+z)^2}\la-\frac{z^2 (2 z-3)}{64 (1+z)^4}\la^2
-\frac{z^2 \left(z^3-15 z^2+23 z-5\right)}{768 (1+z)^6}\la^3
+\mathcal{O}(\la^4),\\
J^{(2)}_A(z,\la)&=
\frac{z \left(1-4z+13 z^2\right)}{384 (1+z)^4}\la^2
+\frac{z \left(19 z^4-170 z^3+168 z^2-26 z+1\right)}{4608
   (1+z)^6}\la^3+\mathcal{O}(\la^4), \\
J^{(3)}_A(z,\la)&=-\frac{z^2 \left(5 z^3-30 z^2+21 z-4\right)}{1536
   (1+z)^6}\la^3+\mathcal{O}(\la^4).
\end{aligned} 
\label{eq:JA-la}
\end{equation}
One can perform a similar computation for the generating function
of symmetric representation in \eqref{eq:gen-sym}
\begin{equation}
\begin{aligned}
 J_S\Bigl(z,\la,\frac{1}{N}\Bigr)&=\sum_{n=0}^\infty N^{-n}J^{(n)}_S(z,\la)
\end{aligned} 
\end{equation}
where the lower order terms are given by 
\begin{equation}
\begin{aligned}
J^{(0)}_S(z,\la)&=
-\log(1-z)+\frac{z}{8(1-z)^2}\la+\frac{z(1+4z+z^2)}{192(1-z)^4}\la^2
+\frac{z \left(z^4+26 z^3+66 z^2+26 z+1\right)}{9216
   (1-z)^6}\la^3+\mathcal{O}(\la^4), \\
J^{(1)}_S(z,\la)&=
\frac{z^2}{8 (1-z)^2}\la+\frac{z^2 (2 z+3)}{64 (1-z)^4}\la^2
+\frac{z^2 \left(z^3+15 z^2+23 z+5\right)}{768 (1-z)^6}\la^3
+\mathcal{O}(\la^4),\\
J^{(2)}_A(z,\la)&=
\frac{z \left(1+4z+13 z^2\right)}{384 (1-z)^4}\la^2
+\frac{z \left(19 z^4+170 z^3+168 z^2+26 z+1\right)}{4608
   (1-z)^6}\la^3+\mathcal{O}(\la^4), \\
J^{(3)}_A(z,\la)&=\frac{z^2 \left(5 z^3+30 z^2+21 z+4\right)}{1536
   (1-z)^6}\la^3+\mathcal{O}(\la^4).
\end{aligned} 
\label{eq:JS-la}
\end{equation}
From \eqref{eq:JA-la} and \eqref{eq:JS-la}, we observe that
$J_A$ and $J_S$ are related by
\begin{equation}
\begin{aligned}
 J_S\Bigl(z,\la,\frac{1}{N}\Bigr)=-J_A\Bigl(-z,\la,-\frac{1}{N}\Bigr).
\end{aligned} 
\label{eq:SvsA}
\end{equation}
Although we do not have a proof of this relation, we expect
that \eqref{eq:SvsA} holds at least in the small $\la$ and $1/N$ expansion 
to all orders. It would be interesting to understand 
the bulk D-brane interpretation of this relation, if any.

\section{Conclusion \label{sec:conclusion}}
In this paper, we have studied the connected correlator
of 1/2 BPS winding Wilson loops. They are on top of each other
along the same circle.
We found the exact expression of 
the connected correlator at finite $N$,
which is written as a trace of some combination of the matrix
$A(k)$.
We also obtained the analytic expressions of the $1/N$ corrections of these
correlators using the topological recursion of Gaussian matrix model.
We have checked the agreement between the exact finite $N$ result and the
analytic result of topological recursion.
We also wrote down the exact form of the generating function
of 1/2 BPS Wilson loops in the symmetric representations.  

There are several open questions. 
It would be very interesting to study the 
$1/N$ corrections of symmetric Wilson loops using our exact result
at finite $N$ \eqref{eq:gen-sym} along the lines of  \cite{CanazasGaray:2018cpk}
and see if the result of \cite{Chen-Lin:2016kkk} is reproduced.
Also, it would be nice to understand
the bulk D-brane picture of the connected
correlator in the large winding number limit \eqref{eq:ka-lim}.
Lastly, it would be very interesting to study the
Wilson loop labeled by a Young diagram with number of boxes of order $N^2$.
Such a Wilson loop is expected to correspond to a bulk geometry with non-trivial topology, known as the bubbling geometry.
It would be very interesting to study the emergence of non-trivial bulk geometry
and its behavior in the quantum regime 
from the exact result of Wilson loops at finite $N$.

\vskip8mm
\acknowledgments
I would like to thank Wolfgang M\"{u}ck for valuable comments
on the draft of this paper. 
This work  was supported in part by JSPS KAKENHI Grant Number 16K05316.



\begin{thebibliography}{99}

\bibitem{Erickson:2000af} 
  J.~K.~Erickson, G.~W.~Semenoff and K.~Zarembo,
  ``Wilson loops in N=4 supersymmetric Yang-Mills theory,''
  \href{http://dx.doi.org/10.1016/S0550-3213(00)00300-X}{Nucl.\ Phys.\ B {\bf 582}, 155 (2000)},
 \href{http://arxiv.org/abs/hep-th/0003055}{[hep-th/0003055]}.

\bibitem{Drukker:2000rr} 
  N.~Drukker and D.~J.~Gross,
  ``An Exact prediction of N=4 SUSYM theory for string theory,''
  \href{http://dx.doi.org/10.1063/1.1372177}{J.\ Math.\ Phys.\  {\bf 42}, 2896 (2001)},
 \href{http://arxiv.org/abs/hep-th/0010274}{[hep-th/0010274]}.


\bibitem{Pestun:2007rz} 
  V.~Pestun,
  ``Localization of gauge theory on a four-sphere and supersymmetric Wilson loops,''
  \href{http://dx.doi.org/10.1007/s00220-012-1485-0}{Commun.\ Math.\ Phys.\  {\bf 313}, 71 (2012)},
 \href{http://arxiv.org/abs/0712.2824}{[arXiv:0712.2824 [hep-th]]}.

\bibitem{Maldacena:1998im} 
  J.~M.~Maldacena,
  ``Wilson loops in large N field theories,''
  \href{http://dx.doi.org/10.1103/PhysRevLett.80.4859}{Phys.\ Rev.\ Lett.\  {\bf 80}, 4859 (1998)},
  \href{http://arxiv.org/abs/hep-th/9803002}{[hep-th/9803002]}.

\bibitem{Rey:1998ik} 
  S.~J.~Rey and J.~T.~Yee,
  ``Macroscopic strings as heavy quarks in large N gauge theory and anti-de Sitter supergravity,''
  \href{http://dx.doi.org/10.1007/s100520100799}{Eur.\ Phys.\ J.\ C {\bf 22}, 379 (2001)},
  \href{http://arxiv.org/abs/hep-th/9803001}{[hep-th/9803001]}.

\bibitem{Drukker:2005kx} 
  N.~Drukker and B.~Fiol,
  ``All-genus calculation of Wilson loops using D-branes,''
  \href{http://dx.doi.org/10.1088/1126-6708/2005/02/010}{JHEP {\bf 0502}, 010 (2005)},
 \href{http://arxiv.org/abs/hep-th/0501109}{[hep-th/0501109]}.

\bibitem{Yamaguchi:2006tq} 
  S.~Yamaguchi,
  ``Wilson loops of anti-symmetric representation and D5-branes,''
  \href{http://dx.doi.org/10.1088/1126-6708/2006/05/037}{JHEP {\bf 0605}, 037 (2006)},
 \href{http://arxiv.org/abs/hep-th/0603208}{[hep-th/0603208]}.

\bibitem{Hartnoll:2006is} 
  S.~A.~Hartnoll and S.~P.~Kumar,
  ``Higher rank Wilson loops from a matrix model,''
  \href{http://dx.doi.org/10.1088/1126-6708/2006/08/026}{JHEP {\bf 0608}, 026 (2006)},
  \href{http://arxiv.org/abs/hep-th/0605027}{[hep-th/0605027]}.

\bibitem{Okuyama:2006jc} 
  K.~Okuyama and G.~W.~Semenoff,
  ``Wilson loops in N=4 SYM and fermion droplets,''
  \href{http://dx.doi.org/10.1088/1126-6708/2006/06/057}{JHEP {\bf 0606}, 057 (2006)},
  \href{http://arxiv.org/abs/hep-th/0604209}{[hep-th/0604209]}.

\bibitem{Gomis:2006sb} 
  J.~Gomis and F.~Passerini,
  ``Holographic Wilson Loops,''
  \href{http://dx.doi.org/10.1088/1126-6708/2006/08/074}{JHEP {\bf 0608}, 074 (2006)},
  \href{http://arxiv.org/abs/hep-th/0604007}{[hep-th/0604007]}.

\bibitem{Gomis:2006im} 
  J.~Gomis and F.~Passerini,
  ``Wilson Loops as D3-Branes,''
  \href{http://dx.doi.org/10.1088/1126-6708/2007/01/097}{JHEP {\bf 0701}, 097 (2007)},
  \href{http://arxiv.org/abs/hep-th/0612022}{[hep-th/0612022]}.

\bibitem{Zarembo:2016bbk} 
  K.~Zarembo,
  ``Localization and AdS/CFT Correspondence,''
\href{http://dx.doi.org/10.1088/1751-8121/aa585b}{J.\ Phys.\ A {\bf 50}, no. 44, 443011 (2017)},
  \href{http://arxiv.org/abs/1608.02963}{[arXiv:1608.02963 [hep-th]]}.

\bibitem{Faraggi:2011ge} 
  A.~Faraggi, W.~M\"{u}ck and L.~A.~Pando Zayas,
  ``One-loop Effective Action of the Holographic Antisymmetric Wilson Loop,''
  \href{http://dx.doi.org/10.1103/PhysRevD.85.106015}{Phys.\ Rev.\ D {\bf 85}, 106015 (2012)},
  \href{http://arxiv.org/abs/1112.5028}{[arXiv:1112.5028 [hep-th]]}.

\bibitem{Faraggi:2011bb} 
  A.~Faraggi and L.~A.~Pando Zayas,
  ``The Spectrum of Excitations of Holographic Wilson Loops,''
  \href{http://dx.doi.org/10.1007/JHEP05(2011)018}{JHEP {\bf 1105}, 018 (2011)},
  \href{http://arxiv.org/abs/1101.5145}{[arXiv:1101.5145 [hep-th]]}.

\bibitem{Faraggi:2014tna} 
  A.~Faraggi, J.~T.~Liu, L.~A.~Pando Zayas and G.~Zhang,
  ``One-loop structure of higher rank Wilson loops in AdS/CFT,''
\href{http://dx.doi.org/10.1016/j.physletb.2014.11.060}{Phys.\ Lett.\ B {\bf 740}, 218 (2015)},
  \href{http://arxiv.org/abs/1409.3187}{[arXiv:1409.3187 [hep-th]]}.

\bibitem{Buchbinder:2014nia} 
  E.~I.~Buchbinder and A.~A.~Tseytlin,
  ``1/N correction in the D3-brane description of a circular Wilson loop at strong coupling,''
\href{http://dx.doi.org/10.1103/PhysRevD.89.126008}{Phys.\ Rev.\ D {\bf 89}, no. 12, 126008 (2014)},
  \href{http://arxiv.org/abs/1404.4952}{[arXiv:1404.4952 [hep-th]]}.

\bibitem{Chen-Lin:2016kkk} 
  X.~Chen-Lin,
  ``Symmetric Wilson Loops beyond leading order,''
  \href{http://dx.doi.org/10.21468/SciPostPhys.1.2.013}{SciPost Phys.\  {\bf 1}, no. 2, 013 (2016)},
  \href{http://arxiv.org/abs/1610.02914}{[arXiv:1610.02914 [hep-th]]}.

\bibitem{Gordon:2017dvy} 
  J.~Gordon,
 ``Antisymmetric Wilson loops in $\mathcal{N}=4$ SYM beyond the planar limit,''
\href{http://dx.doi.org/10.1007/JHEP01(2018)107}{JHEP {\bf 1801}, 107 (2018)},
 \href{http://arxiv.org/abs/1708.05778}{[arXiv:1708.05778 [hep-th]]}.


\bibitem{Yamaguchi:2006te} 
  S.~Yamaguchi,
  ``Bubbling geometries for half BPS Wilson lines,''
\href{http://dx.doi.org/10.1142/S0217751X07035070}{Int.\ J.\ Mod.\ Phys.\ A {\bf 22}, 1353 (2007)},
  \href{http://arxiv.org/abs/hep-th/0601089}{[hep-th/0601089]}.

\bibitem{Lunin:2006xr} 
  O.~Lunin,
  ``On gravitational description of Wilson lines,''
  \href{http://dx.doi.org/10.1088/1126-6708/2006/06/026}{JHEP {\bf 0606}, 026 (2006)},
  \href{http://arxiv.org/abs/hep-th/0604133}{[hep-th/0604133]}.

\bibitem{DHoker:2007mci} 
  E.~D'Hoker, J.~Estes and M.~Gutperle,
  ``Gravity duals of half-BPS Wilson loops,''
  \href{http://dx.doi.org/10.1088/1126-6708/2007/06/063}{JHEP {\bf 0706}, 063 (2007)},
  \href{http://arxiv.org/abs/0705.1004}{[arXiv:0705.1004 [hep-th]]}.

\bibitem{Okuda:2008px} 
  T.~Okuda and D.~Trancanelli,
  ``Spectral curves, emergent geometry, and bubbling solutions for Wilson loops,''
  \href{http://dx.doi.org/10.1088/1126-6708/2008/09/050}{JHEP {\bf 0809}, 050 (2008)},
  \href{http://arxiv.org/abs/0806.4191}{[arXiv:0806.4191 [hep-th]]}.


\bibitem{Fiol:2013hna} 
  B.~Fiol and G.~Torrents,
  ``Exact results for Wilson loops in arbitrary representations,''
 \href{http://dx.doi.org/10.1007/JHEP01(2014)020}{JHEP {\bf 1401}, 020 (2014)},
  \href{http://arxiv.org/abs/1311.2058}{[arXiv:1311.2058 [hep-th]]}.

\bibitem{Horikoshi:2016hds} 
  M.~Horikoshi and K.~Okuyama,
  ``$\alpha'$-expansion of Anti-Symmetric Wilson Loops in $\mathcal{N}=4$ SYM from Fermi Gas,''
  \href{http://dx.doi.org/10.1093/ptep/ptw156}{PTEP {\bf 2016}, no. 11, 113B05 (2016)},
 \href{http://arxiv.org/abs/1607.01498}{[arXiv:1607.01498 [hep-th]]}.

\bibitem{Okuyama:2017feo} 
  K.~Okuyama,
  ``Phase Transition of Anti-Symmetric Wilson Loops in $\mathcal{N}=4$ SYM,''
  \href{http://dx.doi.org/10.1007/JHEP12(2017)125}{JHEP {\bf 1712}, 125 (2017)},
 \href{http://arxiv.org/abs/1709.04166}{[arXiv:1709.04166 [hep-th]]}.

\bibitem{CanazasGaray:2018cpk} 
  A.~F.~Canazas Garay, A.~Faraggi and W.~M\"{u}ck,
  ``Antisymmetric Wilson loops in $N$ = 4 SYM: from exact results to non-planar corrections,''
 \href{http://dx.doi.org/10.1007/JHEP08(2018)149}{JHEP {\bf 1808}, 149 (2018)},
 \href{http://arxiv.org/abs/1807.04052}{[arXiv:1807.04052 [hep-th]]}.

\bibitem{Olesen:2000ji} 
  P.~Olesen and K.~Zarembo,
  ``Phase transition in Wilson loop correlator from AdS / CFT correspondence,''
 \href{http://arxiv.org/abs/hep-th/0009210}{[hep-th/0009210]}.

\bibitem{Gross:1998gk} 
  D.~J.~Gross and H.~Ooguri,
  ``Aspects of large N gauge theory dynamics as seen by string theory,''
  \href{http://dx.doi.org/10.1103/PhysRevD.58.106002}{Phys.\ Rev.\ D {\bf 58}, 106002 (1998)},
 \href{http://arxiv.org/abs/hep-th/9805129}{[hep-th/9805129]}.

\bibitem{Akemann:2001st} 
  G.~Akemann and P.~H.~Damgaard,
  ``Wilson loops in $N$=4 supersymmetric Yang-Mills theory from random matrix theory,''
 \href{http://dx.doi.org/10.1016/S0370-2693(01)00675-X}{Phys.\ Lett.\ B {\bf 513}, 179 (2001)}, 
\href{http://dx.doi.org/10.1016/S0370-2693(01)01346-6}{Erratum: [Phys.\ Lett.\ B {\bf 524}, 400 (2002)]},
\href{http://arxiv.org/abs/hep-th/0101225}{[hep-th/0101225]}.

\bibitem{Kawamoto:2008gp} 
  S.~Kawamoto, T.~Kuroki and A.~Miwa,
  ``Boundary condition for D-brane from Wilson loop, and gravitational interpretation of eigenvalue in matrix model in AdS/CFT correspondence,''
  \href{http://dx.doi.org/10.1103/PhysRevD.79.126010}{Phys.\ Rev.\ D {\bf 79}, 126010 (2009)},
 \href{http://arxiv.org/abs/0812.4229}{[arXiv:0812.4229 [hep-th]]}.

\bibitem{Eynard:2004mh} 
  B.~Eynard,
  ``Topological expansion for the 1-Hermitian matrix model correlation functions,''
  \href{http://dx.doi.org/10.1088/1126-6708/2004/11/031}{JHEP {\bf 0411}, 031 (2004)},
  \href{http://arxiv.org/abs/hep-th/0407261}{[hep-th/0407261]}.

\bibitem{Eynard:2008we} 
  B.~Eynard and N.~Orantin,
  ``Algebraic methods in random matrices and enumerative geometry,''
  \href{http://arxiv.org/abs/0811.3531}{[arXiv:0811.3531 [math-ph]]}.

\bibitem{Haagerup}
U.~ Haagerup and S.~ Thorbj{\o}rnsen,
``Asymptotic expansions for the Gaussian Unitary Ensemble,''
\href{https://doi.org/10.1142/S0219025712500038}{Infinite Dimensional Analysis, Quantum Probability and Related Topics {\bf 15}, 1250003 (2012)},
\href{http://arxiv.org/abs/1004.3479}{[arXiv:1004.3479]}. 

\bibitem{Giombi:2009ms} 
  S.~Giombi, V.~Pestun and R.~Ricci,
  ``Notes on supersymmetric Wilson loops on a two-sphere,''
 \href{http://dx.doi.org/10.1007/JHEP07(2010)088}{JHEP {\bf 1007}, 088 (2010)},
 \href{http://arxiv.org/abs/0905.0665}{[arXiv:0905.0665 [hep-th]]}.



 \end{thebibliography}
\end{document}